\documentclass{article}
\usepackage{float}

\def \lket {|}
\def \rket {\rangle}
\def \lbra {\langle}

\newcommand{\ket}[1]{\lket #1\rket}

\newcommand{\comment}[1]{}
\def\M {{\cal M}}
\def\bbbz{Z}
\newtheorem{Theorem}{Theorem}
\newtheorem{Lemma}{Lemma}
\newtheorem{Claim}{Claim}
\newcommand{\proof}{\noindent {\bf Proof: }}
\newcommand{\qed}{\nobreak \ifvmode \relax \else
      \ifdim\lastskip<1.5em \hskip-\lastskip
      \hskip1.5em plus0em minus0.5em \fi \nobreak
      \vrule height0.75em width0.5em depth0.25em\fi}

\begin{document}
\title{Limits on entropic uncertainty relations for 3 and more MUBs}
\author{Andris Ambainis\thanks{Faculty of Computing, University of Latvia,
Raina bulv. 19, Riga, LV-1586, Latvia, {\tt ambainis@lu.lv}. Supported by University of Latvia
Research Grant ZB01-100 and Marie Curie International Reintegration Grant (IRG).}}
\date{}
\maketitle

\abstract{We consider entropic uncertainty relations for outcomes of
the measurements of a quantum state in 3 or more mutually unbiased
bases (MUBs), chosen from the standard construction of MUBs in prime dimension.
We show that, for any choice of 3 MUBs and at least one choice of
a larger number of MUBs, the best possible entropic uncertainty relation
can be only marginally better than the one that trivially follows from 
the relation by Maassen and Uffink (PRL, 1987) for 2 bases.
}

\section{Introduction}

Uncertainty relations quantify the amount of uncertainty in the outcomes of 
quantum measurements. The most famous uncertainty relation is due to Heisenberg \cite{Heisenberg}
who showed that either the position or the momentum of the particle has at least a 
certain amount of uncertainty.

For finite-dimensional state spaces,
the uncertainty relations are most often stated in terms of the 
entropy of the measurement outcomes \cite{BM,Deutsch,MU,Majernik,WW}. 
Entropic uncertainty relations
have several applications, from locking information in quantum states \cite{D+} to
quantum cryptography in the bounded-storage model \cite{Damgaard+}.
(For more details on those, we refer the reader to the survey by Wehner and Winter \cite{WW}.)

Let $P_1, P_2$ be the probability distributions obtained by measurements
with respect to two orthonormal bases $\M_1, \M_2$ and let $c$ be the maximum of
$|\lbra \psi_1|\psi_2\rket|$, over all $\ket{\psi_1}$ from $\M_1$ 
and $\ket{\psi_2}$ from $\M_2$. Then, as shown by Maasen and Uffink\cite{MU},
\begin{equation}
\label{eq:2ent} 
H(P_1)+H(P_2) \geq -2 \log c .
\end{equation}
The lower bound is maximized if $\M_1$ and $\M_2$ are mutually unbiased. 
Then $|\lbra \psi_1|\psi_2\rket|=\frac{1}{\sqrt{N}}$
and we get a lower bound of $\log N$ on the sum of the two entropies.
This bound is optimal: if we measure a state $\ket{\psi}$ from one of the bases,
the outcome has an entropy of 0 in that basis and an entropy of $\log N$ in the other basis.

In contrast, when we try to quantify the sum of entropies for three or more bases,
fairly little is known. 
Most of the research on this subject 
considers the case when each two of the measurement bases are
mutually unbiased. 

There are two known constructions of mutually unbiased bases (MUBs). 
The first and the most commonly used construction 
is based on generalized Pauli matrices \cite{WF,B+}. 
The second construction \cite{WB} is based on Latin squares.

For either of those constructions, we trivially have
\begin{equation}
\label{eq:trivial} 
H(P_1)+ \ldots + H(P_k) \geq \frac{k}{2} \log N ,
\end{equation}
which follows from dividing the bases into pairs and applying (\ref{eq:2ent}) to each pair.
Better bounds are known for the case when the number of measurement bases is large (i.e. we use 
the full collection of $d+1$ MUBs in dimension $d$ or a large subset of it 
\cite{Ivanovic,Sanchez,SanchezR}). 
But, for the case when we consider a small number of measurements, 
only two partial results are known, one for each of the two constructions of MUBs.

For the first construction, computer simulations by DiVincenzo et al. \cite{D+} 
(for the number of bases $k$ from 3 to 29) indicate
\[ H(P_1)+ \ldots + H(P_k) \approx ck\log N ,\]
with $c$ scaling as $1-\epsilon - \frac{1}{k}$ where $\epsilon$ is between 0.10 and 0.15. 
For the second construction, Ballester and Wehner \cite{BW} show that (\ref{eq:trivial}) is essentially
optimal and no better bound can be achieved, even when we use the maximum number of 
MUBs provided by the Latin square construction.

Thus, it seems that the two constructions display a significantly different behaviour: one 
provides better and better uncertainty relations as we increase the number of bases (which is good
for applications such as locking of correlations in quantum states \cite{D+}) while the second does not.

In this note, we provide some new results which show that the first constructions of MUBs also 
fails to give better uncertainty relations in some situations:
\begin{enumerate}
\item
For any 3 bases from this construction, we can find a state $\ket{\psi}$ with
\[ H(P_1)+ H(P_2) + H(P_3) \leq \left( \frac{3}{2}+o(1) \right) \log N  .
\]
Thus, the trivial bound (\ref{eq:trivial}) is nearly tight in this case.
\item
For any $k\leq n^{\epsilon}$, we can select $k$ MUBs in dimension $n$ so that
\[ H(P_1)+ \ldots + H(P_k) \leq (1+\epsilon+o(1))\frac{k}{2} \log N .\]
\end{enumerate} 
Our results do not rule out the possibility of 
good uncertainty relations for $k\geq 4$ MUBs but indicate
that a careful choice of the set of MUBs may be necessary to obtain such relations.

\section{MUBs in prime dimensions}
\label{sec:bg}

In this section, we first describe the Wootters-Fields \cite{WF} construction
of mutually unbiased bases and then analyze its symmetry properties.
The results on symmetry properties will be used to prove our bound
on entropic uncertainty relations for 3 MUBs in section \ref{sec:3mubs}.

The Wootters and Fields \cite{WF} construction for prime dimension $p$
is as follows.
The first MUB, $\M_c$, just consists of the computational 
basis states $\ket{0}$, $\ket{1}$, $\ldots$, $\ket{p-1}$.
The other $p$ MUBs are denoted $\M_0$, $\ldots$, $\M_{p-1}$,
with $\M_j$ consisting of states 
$\ket{\psi_{j, 0}}$, $\ldots$, $\ket{\psi_{j, p-1}}$
defined by
\[ \ket{\psi_{j, k}}=\sum_{l\in \{0, 1, \ldots, p-1\}}
 w^{j\cdot l^2+k\cdot l} \ket{l} \]
where $w=e^{2\pi i/p}$.

We say that two triplets of MUBs ($\M'_1$, $\M'_2$, $M'_3$) and 
($\M''_1$, $\M''_2$, $\M'''_3$) are equivalent if there is unitary $U$
that maps $\M'_1$, $\M'_2$, $M'_3$ to $\M''_1$, $\M''_2$, $\M'''_3$
(in some order).

\begin{Lemma}
\label{claim:2bases}
Let $d$ be the smallest quadratic nonresidue $\bmod~p$.
Any set of 3 different MUBs selected from $\M_c$, $\M_0$, $\ldots$, $\M_{p-1}$
is equivalent to either the set $\M_c$, $\M_0$, $M_{1}$ 
or the set $\M_c$, $\M_0$, $\M_d$.
\end{Lemma}

\proof
Let $\M'_1, \M'_2, \M'_3$ be three different MUBs (selected from $\M_c$, 
$\M_0$, $\ldots$, $\M_{p-1}$).
We consider the following two unitary transformations that permute the MUBs:
\begin{itemize}
\item
The unitary transformation $W\ket{l}= w^{l^2} \ket{l}$
leaves $\M_c$ unchanged and maps $\M_j$ ($j\in\{0, 1, \ldots, p-1\}$) 
to $\M_{(j+1) \bmod p}$.
\item
Quantum Fourier transform 
\[ F\ket{l}=\sum_{j=0}^{p-1} \frac{1}{\sqrt{p}} w^{j\cdot l} \ket{j} \]
maps $\M_c$ and $\M_0$ one to another and permutes $\M_1, \ldots, \M_{p-1}$ in
some way.
\end{itemize}
We can map $\M'_1$, $\M'_2$, $\M'_3$ 
to $\M_c$, $\M_0$ and $\M_k$ (for some $k\in\{0, 1, \ldots, p-1\}$) as follows:
\begin{enumerate}
\item
We repeatedly apply $W$ until one of $\M'_i$ is mapped to $\M_0$.
\item
We then apply $F$, mapping $\M_0$ to $\M_c$.
\item
We then repeatedly apply $W$ until one of the other MUBs is mapped to $\M_0$.
\end{enumerate}

Next, let $a\in\{1, \ldots, p-1\}$.
Define $U_a\ket{j}=\ket{(a^{-1}j)\bmod p}$ (where $a^{-1}$ is the inverse of
$a$ in $\bbbz_p$).
The transformation $U_a$ leaves $\M_c$ unchanged (permuting the basis states in this basis). 
For the basis $\M_j$, we have
\[ U_a \ket{\psi_{j, k}} = \sum_{l\in \{0, 1, \ldots, p-1\}}
w^{j\cdot l^2+k\cdot l} \ket{(a^{-1} l) \bmod p} = \]
\[ \sum_{l \in\{0, 1, \ldots, p-1\}} 
w^{j\cdot a^2 l^2+k} \ket{l} =
\ket{\psi_{(a^2 j) \bmod p, (ak)\bmod p}} .\]
Thus, the basis $\M_j$ is mapped to $M_{(a^2 j) \bmod p}$.
In particular, this means that $\M_0$ is mapped to itself.

If we have a set $\M_c$, $\M_0$, $\M_j$ with $j$ being a quadratic residue,
then $j^{-1}$ is a quadratic residue as well.
Let $a$ be a solution of $x^2\equiv j^{-1} (\bmod p)$.
Then, $U_a$ leaves $\M_c$ and $\M_0$ unchanged and maps
$M_j$ to $M_{(a^2 j) \bmod p}=M_1$.

If we have a set $\M_c$, $\M_0$, $\M_j$ with $j$ being a quadratic nonresidue,
then $j^{-1}$ is a quadratic non-residue and $j^{-1} d$ is a quadratic residue
(modulo a prime, a product of two quadratic non-residues is a quadratic residue).
Let $a$ be a solution of $x^2\equiv j^{-1} d (\bmod~p)$.
Then, $U_a$ leaves $\M_c$ and $\M_0$ unchanged and maps
$M_j$ to $M_{(a^2 j) \bmod p}=M_d$.
\qed

\begin{Lemma}
The sets of MUBs $\M_c$, $\M_0$, $\M_1$ and $\M_c$, $\M_0$, $\M_d$ are
equivalent if and only if the prime $p$ is of the form $p=4k+3$, $k\in\bbbz$.
\end{Lemma}

\proof
If $p=4k+3$, then -1 is a quadratic non-residue $\bmod~p$ \cite{W}.
As shown in the proof of Lemma \ref{claim:2bases}, $\M_c$, $\M_0$, $\M_d$
is then equivalent to $\M_c$, $\M_0$, $\M_{-1}$.
Applying the unitary transformation
$W$ from Lemma \ref{claim:2bases}
maps $\M_c, \M_0, \M_{-1}$ to $\M_c$, $\M_{1}$ and $\M_0$.

Next, we consider the case when $p=4k+1$. 
Then, -1 is a quadratic residue \cite{W}. We first show

\begin{Claim}
Assume that -1 is a quadratic residue $\bmod~p$.
Then, any permutation of $\M_c, \M_0$ and $\M_1$ can be implemented by
a unitary transformation.
\end{Claim}

\proof
It suffices to show that we can implement the following two permutations:
\[ \M_c \rightarrow \M_0, \M_0 \rightarrow \M_c, \M_1 \rightarrow \M_1 \]
\[ \M_c \rightarrow \M_c, \M_0 \rightarrow \M_1, \M_1 \rightarrow \M_0 \]
because any permutation can be expressed as a product of those.
Those two transformations can be implemented as follows:
\begin{enumerate}
\item
The quantum Fourier transform $F$ transforms bases in a following way: 
$F(\M_c)=\M_0$, $F(\M_0)=\M_c$, $F(\M_1)=\M_a$ where $a$
is the unique element of $\{0, 1, \ldots, p-1\}$ satisfying
$4a\equiv -1 (\bmod~p)$.
We can then transform these bases to $\M_0$, $\M_c$, $\M_{-1}$ by
applying the transformation $U_2$ defined in the proof of Lemma 
\ref{claim:2bases}.

Since $-1$ is a quadratic residue mod $p$, there exists $x$ such that
$x^2\equiv -1 (\bmod~p)$. Applying $U_x$ maps $\M_0, \M_c, \M_{-1}$
to $\M_0, \M_c, \M_1$.
\item
We first apply $U_x$ mapping $\M_c, \M_0, \M_{1}$
to $\M_c, \M_0, \M_{-1}$.
We then apply $U\ket{l}= e^{2\pi i \frac{l^2}{p}} \ket{l}$
which maps those to $\M_c, \M_1, \M_0$.
\end{enumerate}
\qed

Therefore, if we have a unitary transformation $U$ that transforms 
$\M_c, \M_0, \M_1$ to $\M_c, \M_0, \M_d$, we can assume
that it implements the following map
\[ \M_c \rightarrow \M_c, \M_0 \rightarrow \M_0, \M_1 \rightarrow \M_d .\]
Since $U$ fixes $\M_c$, $U$ is of the form 
\[ U\ket{i}=\lambda(i) \ket{f(i)}, \] 
where $f(0), \ldots, f(p-1)$
is a permutation of $0, \ldots, p-1$ and $\lambda(i)$ are complex numbers of absolute value 1.
Without a loss of generality, we can assume that $\lambda(0)=1$.
The other $\lambda(i)$ all must be powers of $w$
(otherwise, vectors in $\M_0$ (whose coefficients
are powers of $w$) would not be mapped to vectors in $\M_0$).
Let
\begin{equation}
\label{eq:01} U\ket{0}=\ket{i_0} , U\ket{1}=w^{k_1} \ket{i_1} .
\end{equation}
We claim that this implies 
\begin{equation}
\label{eq:constrained} 
U\ket{j}=w^{k_1 j} \ket{i_0+j(i_1-i_0)} .
\end{equation}

To show that, we first assume 
\[ U\ket{j}=w^{k_j} \ket{i_j} .\]
We consider the state 
$\ket{\psi_{0, 0}}=\sum_{i=0}^{p-1} \frac{1}{\sqrt{d}} \ket{i}$
which belongs to the basis $\M_0$.
It must get mapped to a state in $\M_0$ and
the only possibility that is consistent with (\ref{eq:01}) is
\[U\ket{\psi_{0, 0}} = w^{-k i_0} \ket{\psi_{0, k}} 
\]
where $k= \frac{k_1}{i_1-i_0}$ (with all operations modulo $p$).
Then, we must have
\begin{equation}
\label{eq:choice1} 
 U\ket{j}=w^{k (i_j-i_0)} \ket{i_j} . 
\end{equation}
Similarly, the state 
$\ket{\psi_{0, 1}}=\sum_{i=0}^{p-1} \frac{w^i}{\sqrt{d}} \ket{i}$
must also get mapped to a state in $\M_0$ and the only possibility
consistent with (\ref{eq:01}) is
\[ U\ket{\psi_{0, 1}} = w^{-k' i_0} \ket{\psi_{0, k'}} .
\]
where $k'=\frac{k_1+1}{i_1-i_0}$.
Then, we must have
\begin{equation}
\label{eq:choice2} 
 U\ket{j} = w^{k'(i_j-i_0) - j} \ket{i_j} .
\end{equation}
Since $i_j$ must have the same coefficients in (\ref{eq:choice1}) and (\ref{eq:choice2}),
we have
\[ \frac{k_1}{i_1-i_0} (i_j - i_0) = \frac{k_1+1}{i_1-i_0} (i_j - i_0) - j \]
and
\[ j = \frac{i_j-i_0}{i_1-i_0} \]
which is equivalent to $i_j=i_0+j(i_1-i_0)$.
The coefficient of $\ket{i_j}$ in (\ref{eq:choice1}) is 
\[ w^{k (i_j-i_0)} = w^{\frac{k_1}{i_1-i_0} (i_j - i_0)} = w^{k_1 j} .\]
This implies (\ref{eq:constrained}).

Next, a transformation of the form (\ref{eq:constrained}) can be expressed as a 
product of three transformations:
\begin{enumerate}
\item
$\ket{j}\rightarrow w^{b j} \ket{j}$ for some $b\in\{0, 1, \ldots, p-1\}$;
\item
$\ket{j}\rightarrow \ket{cj}$ for some $c\in\{1, \ldots, p-1\}$;
\item
$\ket{j}\rightarrow \ket{j+d}$ for some $d\in\{0, 1, \ldots, p-1\}$.
\end{enumerate}
The second transformation maps $\M_1$ to $\M_{c^2}$. 
The first and the third transformation just permute the vectors within each $\M_i$.
Therefore, we can map $\M_1$ to $\M_{c^2}$ but not to $\M_d$ where $d$ is a quadratic non-residue.
\qed

For our result, we also need an upper bound on the smallest quadratic non-residue
$d$. It is known that:
\begin{itemize}
\item
If $p=8k+5$, $k$-integer, then $d=2$.
\item
For $p=8k+1$, then $d=O(\log^2 p)$ for all $p$, assuming the generalized Riemann hypothesis 
is true \cite{W}.
\item
Although $d$ is small for most primes $p$, no good bound without the use of GRH is known
\cite{Gowers,Tao} 
\end{itemize}

\section{Limit on entropic uncertainty relations}

\subsection{Measurement in 3 bases}
\label{sec:3mubs}

As shown in the previous section, any set of 3 MUBs is equivalent
to $\M_c, \M_0, \M_1$ or $\M_c, \M_0, \M_d$.
We first consider the case of measurements $\M_c, \M_0, \M_1$.

\begin{Theorem}
\label{thm:unc3}
Let $E(\psi)$ be the average of the entropies of probability distributions
obtained by measuring $\psi$ in the bases $\M_c, \M_0, \M_1$.
There exists a state $\ket{\psi}$ such that
\[ E(\psi) \leq \frac{1}{2} \log p +\frac{1}{6} \log \log p + c \]
for some constant $c$.
\end{Theorem}

\proof
Let $\ket{\psi}=\frac{1}{\sqrt{m}}\sum_{j=0}^{m-1} \ket{j}$, where
\begin{equation}
\label{eq:mdef} 
m = \left\lfloor \sqrt{\frac{p}{4 \pi \log p}} \right\rfloor .
\end{equation}
Measuring $\ket{\psi}$ in $\M_c$ produces one of values 
$0, 1, \ldots, m-1$ with probability $\frac{1}{m}$ each.
This probability distribution has the entropy of $\log m$.

\begin{Lemma}
\label{lem:ent1}
Measuring $\ket{\psi}$ in $\M_0$ produces a probability distribution with an
entropy of at most $\log p-\log m+10$.
\end{Lemma}
 
\proof
Let $|k|=\min(k, p-k)$.
Measuring $\ket{\psi}$ in $\M_0$ gives the value $k$ with probability
\[
\left| \frac{1}{\sqrt{m p}} \sum_{j=0}^{m-1} 
e^{2\pi i \frac{jk}{p}} \right|^2 =
\frac{1}{mp} \left| \frac{e^{2\pi i \frac{km}{p}}-1}{e^{2\pi i \frac{k}{p}}-1} \right|^2 
\]
\begin{equation}
\label{eq:prob} 
\leq \frac{4}{mp}  \frac{1}{\left|e^{2\pi i \frac{k}{p}}-1\right|^2}
\leq \frac{\pi^2 p}{|k|^2 m} ,
\end{equation}
where the last inequality
follows from $|e^{ix}-1|\geq \frac{2|x|}{\pi}$
being true for all $x\in[-\pi, \pi]$.

Let $t=\lceil \frac{8\pi^2 p}{m} \rceil$.
Let $S$ be the set of all $k$ with $|k|< t$
and let $S_i$ (for $i=0, 1, \ldots$)
be the set of all $k$ with $2^i t \leq
|k| < 2^{i+1} t$.

\begin{Claim}
Let $p_i$ be the probability of measuring $k\in S_i$.
Then,
\[ p_i \leq \frac{1}{2^{i+2}} .\]
\end{Claim}

\proof
If $|k|\geq 2^{i} t$, the probability (\ref{eq:prob}) is at most
$\frac{\pi^2 p}{2^{2i} t^2 m}$. 
Since there are $2^{i+1} t$ values $k\in S_i$, we have 
\[ p_i \leq 2^{i+1} t \frac{\pi^2 p}{2^{2i} t^2 m} =
\frac{\pi^2 p}{2^{i-1} t m} \leq \frac{1}{4} ,\]
with the last inequality following from the definition of $t$.
\qed

This claim also implies that
\[ \sum_{i} p_i \leq \sum_{i=0}^{\infty} \frac{1}{2^{i+2}} = \frac{1}{2}.\]

The entropy of the probability distribution of outcomes of $\M_0$ is
upper-bounded by the entropy of the probability distribution in which 
each element of $S_i$ has a probability $\frac{p_i}{|S_i|}$ and
each element of $S$ has a probability $\frac{p_0}{|S|}$,
where $p_0=1-\sum_{i\geq i_0} p_i$.
The entropy of this probability distribution is 
\[ -|S| \frac{p_0}{|S|} \log \frac{p_0}{|S|} - 
\sum_{i\geq i_0} |S_i| \frac{p_i}{|S_i|} \log \frac{p_i}{|S_i|} = \]
\[ -p_0 \log \frac{p_0}{|S|} -
\sum_{i\geq i_0} p_i \log \frac{p_i}{|S_i|} = \]
\[ (p_0 \log |S| + \sum_{i\geq i_0} p_i \log |S_i|) -
(p_0 \log p_0 + \sum_{i\geq i_0} p_i \log p_i) .\]
Since $|S| \leq 2t$ and $|S_i|\leq 2^{i+1} t$, we can upperbound
the first component by
\[ p_0 (1+\log t) + \sum_{i\geq 0} p_i (i+1+\log t) 
= (1+\log t) + \sum_{i\geq 0} p_i i \]
\[ 
\leq (1+\log t) + \sum_{i\geq 0} \frac{1}{2^{i+2}} i 
\leq \log t + \frac{3}{2} .\]
For the second component, we have 
\[ - (p_0 \log p_0 + \sum_{i\geq i_0} p_i \log p_i) \leq 
- \frac{1}{2} \log \frac{1}{2} - \sum_{i\geq 2} \frac{1}{2^i} \log \frac{1}{2^i}
\leq \frac{1}{2} + \sum_{i\geq 2} \frac{1}{2^{i}} i \leq 2 .\]
Therefore, the entropy is at most 
\[ \log t + \frac{7}{2} \leq \log m - \log p + 10 ,\]
with the last inequality following from the definition of $t$. 
\qed 

\begin{Lemma}
\label{lem:ent2}
Assume that $m^2 \leq \frac{p}{4 \pi \log p}$.
Let $H_0$ and $H_1$ be the entropies of the probability distributions
obtained by measuring $\ket{\psi}$ in $\M_0$ and $\M_1$.
Then,
\[ H_1 \leq H_0+1.\]
\end{Lemma}

\proof
Measuring $\ket{\psi}=\frac{1}{\sqrt{m}} \sum_{j=0}^m \ket{j}$ in
the basis $\M_1$ produces the same probability distribution as
measuring $\ket{\psi'}=\frac{1}{\sqrt{m}} \sum_{j=0}^m 
e^{-2\pi i \frac{j^2}{p}} \ket{j}$ in the basis $\M_0$. 
We have 
\[ |e^{-2\pi i \frac{j^2}{p}}-1| \leq 2 \pi \frac{j^2}{p} .\]
Therefore, $\|\psi-\psi'\|\leq 2\pi \max_j \frac{j^2}{p} = 2 \pi \frac{m^2}{p}$. 
The variational distance between the probability distributions
obtained by measuring $\ket{\psi}$ and $\ket{\psi'}$ is
at most $2 \|\psi-\psi'\|\leq 4\pi \frac{m^2}{p}$.
Because of the definition of $m$,
this is at most $\frac{1}{\log  p}$.
Lemma \ref{lem:ent2} now follows from the lemma below.

\begin{Lemma}
\cite{CK}
\label{lem:entdif}
Let $P, P'$ be probability distributions over a $p$ element set and
$|P-P'|\leq \delta$. Then, 
\[ |H(P)-H(P')|\leq H(\delta)+\delta \log(p-1). \]
\end{Lemma}

\qed

By combining all the bounds on the entropies
(the trivial $\log m$ bound on the entropy of the measurement in
$\M_c$, Lemma \ref{lem:ent1} and
Lemma \ref{lem:ent2}), the average of entropies must
be at most
\[ \frac{2}{3} \log p -\frac{1}{3} \log m + 7 + o(1) .\]
Substituting (\ref{eq:mdef}) instead of $m$ completes
the proof of the theorem.
\qed

For the case when the set of 3 MUBs consists of $\M, \M_0$ and $\M_d$,
a similar proof gives
\[ E(\psi) \leq \frac{1}{2} \log p +\frac{1}{6} \log \log p
+\frac{1}{6} \log d+c .\]
(The main difference is that we have to take
\[ m = \left\lfloor \sqrt{\frac{p}{4 \pi d \log p}} \right\rfloor \]
instead of (\ref{eq:mdef}.)
As discussed at the end of section \ref{sec:bg},
generalized Riemann hypothesis (GRH) implies
$d= O(\log^2 p)$ and $\log d \leq 2 \log \log p + O(1)$
for all $p$. Thus, we have 

\begin{Theorem}
Let $\M'_1, \M'_2, \M'_3$ be an arbitrary subset of $\M_c$, $\M_0$, $\ldots$, $\M_{p-1}$.
Let $E(\psi)$ be the average of the entropies of probability distributions
obtained by measuring $\psi$ in $\M'_0, \M'_1, \M'_{2}$.
we have:
\begin{enumerate}
\item
If $p$ is not of the form $p=8k+1$, there exists a state $\ket{\psi}$ such that
\[ E(\psi) \leq \frac{1}{2} \log p +\frac{1}{6} \log \log p + c \]
for some constant $c$.
\item
If $p=8k+1$ and GRH is true, there exists a state $\ket{\psi}$ such that
\[ E(\psi) \leq \frac{1}{2} \log p +\frac{1}{2} \log \log p + c \]
for some constant $c$.
\end{enumerate}
\end{Theorem}

\subsection{Measurement in a larger number of bases}

\begin{Theorem}
Let $E(\psi)$ be the average of the entropies of probability distributions
obtained by measuring $\psi$ in the bases $\M_c, \M_0, \ldots, 
M_{\lfloor p^{\epsilon} \rfloor}$.
There exists state $\ket{\psi}$ such that
\[ E(\psi) \leq \frac{1+\epsilon}{2} \log p + o(\log p) \]
for some constant $c$.
\end{Theorem}

\proof
As in the proof of Theorem \ref{thm:unc3}, we take 
$\ket{\psi}=\frac{1}{\sqrt{m}}\sum_{j=0}^{m-1} \ket{j}$.
But we now choose
\begin{equation}
\label{eq:mdef1} 
m = \left\lfloor \frac{p^{\frac{1-\epsilon}{2}}}{\sqrt{4 \pi \log p}} \right\rfloor .
\end{equation}
Then, the entropy of measuring $\ket{\psi}$ in the basis $\M_c$ is $\log m$ and 
the entropy of measuring of measuring $\ket{\psi}$ in the basis $\M_0$ is $\log p-\log m+10$
(by Lemma \ref{lem:ent1}).

We now bound the entropy of measuring $\ket{\psi}$ in a basis $\M_k$, $k\in\{1, \ldots,
\lfloor n^{\epsilon} \rfloor\}$. 

Similarly to the proof of Lemma \ref{lem:ent2}, measuring 
$\ket{\psi}=\frac{1}{\sqrt{m}} \sum_{j=0}^m \ket{j}$ in
the basis $\M_k$ produces the same probability distribution as
measuring $\ket{\psi'}=\frac{1}{\sqrt{m}} \sum_{j=0}^m 
e^{-2\pi i \frac{k j^2}{p}} \ket{j}$ in the basis $\M_0$. 
We have 
\[ |e^{-2\pi i \frac{k j^2}{p}}-1| \leq 2 \pi \frac{k j^2}{p} .\]
We can upperbound $k$ by its maximum value, $p^{\epsilon}$ and $j$ by its maximum
value, $m$.
By summing over all $j\in\{0, 1, \ldots, m-1\}$, we get
\[ \|\psi-\psi'\|\leq 2\pi \frac{p^{\epsilon} m^2}{p} \leq \frac{1}{2 \log p}. \]
The variational distance between the probability distributions
obtained by measuring $\ket{\psi}$ and $\ket{\psi'}$ is
at most $2 \|\psi-\psi'\|\leq \frac{1}{\log  p}$.
By Lemma \ref{lem:entdif}, this means that the entropies of 
the two probability distributions differ by at most $1+o(1)$.
Since the entropy of the distribution obtained by measuring $\ket{\psi}$
in $\M_0$ is $\log p- \log m +10$, this means that the entropy of the distribution
obtained by measuring $\ket{\psi}$ in $\M_k$ is at most
\[ \log p - \log m +11 + o(1) .\]
By substituting (\ref{eq:mdef1}) instead of $m$, this is at most
\[ \frac{1+\epsilon}{2} \log p + o(\log p) .\]
This upperbounds the entropy for $\M_0, \M_1, 
\ldots, M_{\lfloor p^{\epsilon}\rfloor}$.
For $\M_c$, the entropy is $\log m \leq \frac{1-\epsilon}{2} p$.
Therefore, the theorem follows.
\qed

\end{document}